\documentclass{mn2e}

\usepackage{amsfonts}
\usepackage{amsmath}
\usepackage{graphicx}

\title[The luminosity-weighted correlation function]
      {The luminosity-weighted or `marked' correlation function}
 \author[R. Skibba, R. K. Sheth, A. J. Connolly \& R. Scranton]
 {Ramin Skibba$^1$, Ravi K. Sheth$^2$, Andrew J. Connolly$^1$, \& Ryan Scranton$^1$
  \thanks{E-mail:  ramin@phyast.pitt.edu (RS); 
                   shethrk@physics.upenn.edu (RKS);
                   ajc@phyast.pitt.edu (AJC)
                   scranton@bruno.phyast.pitt.edu (RS)}\\
  $^{1}$Department of Physics \& Astronomy, University of Pittsburgh, 
        Pittsburgh, PA 15260, USA\\
  $^{2}$Department of Physics \& Astronomy, University of Pennsylvania, 
        Philadelphia, PA 19130, USA}

\begin{document}

\pagerange{\pageref{firstpage}--\pageref{lastpage}}

\maketitle
\label{firstpage}

\begin{abstract}
We present measurements of the redshift-space luminosity-weighted 
or `marked' correlation function in the SDSS.  These are compared 
with a model in which the luminosity function and luminosity dependence 
of clustering are the same as that observed, and in which the form of 
the luminosity-weighted correlation function is entirely a consequence 
of the fact that massive halos populate dense regions.  
We do this by using mock catalogs which are constrained to reproduce 
the observed luminosity function and the luminosity dependence of 
clustering, as well as by using the language of the redshift-space 
halo-model.  
These analyses show that marked correlations may show a signal on 
large scales even if there are no large-scale physical effects---the 
statistical correlation between halos and their environment will 
produce 
a measureable signal.  
Our model is in good agreement with the measurements, indicating that 
the halo mass function in dense regions is top-heavy; 
the correlation between halo mass and large scale environment is the 
primary driver for correlations between galaxy properties and 
environment; 
and the luminosity of the central galaxy in a halo is different from 
(in general, brighter than) that of the other objects in the halo.  
Thus our measurement provides strong evidence for the accuracy of these 
three standard assumptions of galaxy formation models.  These 
assumptions 
also form the basis of current halo-model based interpretations of 
galaxy clustering.  

When the same galaxies are weighted by their $u-$, $g-$, or $r-$band 
luminosities, then the marked correlation function is stronger in the 
redder bands.  When the weight is galaxy color rather than luminosity, 
then the data suggest that close pairs of galaxies tend to have redder 
colors.  This wavelength dependence of marked correlations is in 
qualitative agreement with galaxy formation models, and reflects the 
fact that the mean luminosity of galaxies in a halo depends more 
strongly on halo mass in the $r-$band than in $u$.  The $u-$band 
luminosity is a tracer of star formation, so our measurement suggests 
that the correlation between star formation rate and halo mass is 
not monotonic.  In particular, the luminosity and color dependence we 
find are consistent with models in which the galaxy population in 
clusters is more massive and has a lower star formation rate than 
does the population in the field.  The virtue of this measurement of 
environmental trends is that it does not require classification of 
galaxies into field, group and cluster environments.  
\end{abstract}



\section{Introduction}
In hierarchical models of structure formation, there is a correlation 
between halo formation and abundances and the surrounding large scale 
structure---the mass function in dense regions is top-heavy 
(Mo \& White 1996; Sheth \& Tormen 2002).  
Galaxy formation models assume that the properties of a galaxy are 
determined entirely by the mass and formation history of the dark 
matter halo within which it formed.  
Thus, the correlation between halo properties and environment 
induces a correlation between galaxy properties and environment.  
The main goal of the present work is to test if this statistical 
correlation accounts for most of the observed trends between 
luminosity and environment (luminous galaxies are more strongly 
clustered), or if other physical effects also matter.  

We do so by using the statistics of marked correlation functions 
(Stoyan \& Stoyan 1994; Beisbart \& Kerscher 2002) which have 
been shown to provide sensitive probes of environmental effects 
(Sheth \& Tormen 2004; Sheth, Connolly \& Skibba 2005).
The halo model (see Cooray \& Sheth 2002 for a review) is the 
language currently used to interpret measurements of galaxy 
clustering.  Sheth (2005) develops the formalism for including 
marked correlations in the halo model of clustering, and 
Skibba \& Sheth (2005) extend this to describe measurements made 
in redshift space.  This halo model provides an analytic description 
of marked statistics when correlations with environment arise 
entirely because of the statistical effect.  

Section~\ref{mocks} describes how to construct a mock galaxy catalog 
in which the luminosity function and the luminosity dependence of 
clustering are the same as those observed in the Sloan Digital Sky 
Survey.  In these mock catalogs, any correlation 
with environment is {\em entirely} due to the statistical effect.  
Section~\ref{model} shows that the halo model description of marked 
statistics provides a good description of this effect, both in real 
and in redshift space.  
Section~\ref{sdss} compares measurements of marked statistics in the 
SDSS with the halo model prediction.  The comparison provides a test 
of the assumption that correlations with environment arise entirely 
because of the statistical effect.  A final section summarizes our 
results, and shows that marked statistics provide interesting 
information about the correlation between galaxies and their 
environments without having to separate the population into the 
two traditional extremes of `cluster' and `field'.  

\section{Weighted or marked correlations in the `standard' 
model}\label{mocks}
Zehavi et al. (2005) have measured the luminosity dependence of 
clustering in the SDSS (York et al. 2000; Adelman-McCarthy et al. 2005).  
They interpret their measurements using the language of the halo model 
(see Cooray \& Sheth 2002 for a review).  In particular, they describe 
how the distribution of galaxies depends on halo mass in a $\Lambda$CDM 
model with $(\Omega_0,h,\sigma_8) = (0.3,0.7,0.9)$ which is spatially 
flat.  
In this description, only sufficiently massive halos ($M_{\rm 
halo}>10^{11}M_\odot$) host galaxies. 
Each sufficiently massive halo hosts a galaxy at its centre, and may 
host satellite galaxies.  The number of satellites follows a Poisson 
distribution with a mean value which increases with halo mass 
(following Kravtsov et al. 2004).  
In particular, Zehavi et al. report that the mean number of galaxies 
with luminosity greater than $L$ in halos of mass $M$ is 
\begin{equation}
 N_{\rm gal}(>L|M) = 1 + N_{\rm sat}(>L|M) 
               = 1 + \left[{M\over M_1(L)}\right]^{\alpha(L)}
 \label{sdssNg}
\end{equation}
if $M\ge M_{\rm min}(L)$, and $N_{\rm gal}(M)=0$ otherwise.  
In practice, $M_{\rm min}(L)$ is a monotonic function of $L$; 
we have found that their results are quite well approximated by 
\begin{equation}
 \left({M_{\rm min}\over 10^{12}h^{-1}M_\odot}\right)\approx 
    \exp\left({L\over 1.4\times 10^8h^{-2}L_\odot}\right) - 1,  
 \label{MLapprox}
\end{equation}
$M_1(L)\approx 23\,M_{\rm min}(L)$, and $\alpha\sim 1$.  

Later in this paper we will also study a parametrization in which 
the cutoff at $M_{\rm min}$ is less abrupt:
\begin{eqnarray}
 N_{\rm gal}(>L|M) &=& {\rm erfc}
      \left[{\log_{10} M_{\rm min}(L)/M\over \sqrt{2}\sigma}\right] 
      + N_{\rm sat}(>L|M)\nonumber\\
 N_{\rm sat}(>L|M) &=& \left[{M\over M_1(L)}\right]^{\alpha(L)}.
 \label{sdssNerfc}
\end{eqnarray}
This is motivated by the fact that semi-analytic galaxy formation 
models show smoother cut-offs at low-masses (Sheth \& Diaferio 2001; 
Zheng et al. 2005), and that parameterizations like this one can 
also provide good fits to the SDSS measurements (Zehavi et al. 2005).  

We use the model in equation~(\ref{sdssNg}) to populate halos in the 
$z=0.13$ outputs of the VLS $\Lambda$CDM simulation 
(Yoshida, Sheth \& Diaferio 2001) as follows.  
We specify a minimum luminosity $L_{\rm min}$ which is smaller than the 
minimum luminosity we wish to study.  
We then select the subset of halos in the simulations which have 
$M>M_{\rm min}(L_{\rm min})$.  We specify the number of satellites each 
such halo contains by choosing an integer from a Poisson distribution 
with mean $N_{\rm sat}(>L_{\rm min}|M)$.  We then specify the 
luminosity 
of each satellite galaxy by generating a random number $u$ distributed 
uniformly between 0 and 1, and finding that $L$ for which 
 $N_{\rm sat}(>L|M)/N_{\rm sat}(>L_{\rm min}|M) = u$.  
This ensures that the satellites have the correct luminosity 
distribution.  
Finally, we distribute the satellites around the halo centre so that 
they follow an NFW profile (see Scoccimarro \& Sheth 2002 for details).  
We also place a central galaxy at the centre of each halo.  
The luminosity of this central galaxy is given by inverting the 
$M_{\rm min}(L)$ relation between minimum mass and luminosity.  
We assign redshift space coordinates to the mock galaxies by assuming 
that a galaxy's velocity is given by the sum of the velocity of its 
parent halo plus a virial motion contribution which is drawn from a 
Maxwell-Boltzmann distribution with dispersion which depends on halo 
mass (following equation~\ref{sigmavir} below).  
We insure that the centre of mass motion of all the satellites in a 
halo is the same as that of the halo itself by subtracting the mean 
virial motion vector of satellites from the virial motion of each 
satellite (see Sheth \& Diaferio 2001 for tests which indicate that 
this model is accurate).  

\begin{figure}
 \centering
  \includegraphics[width=\hsize]{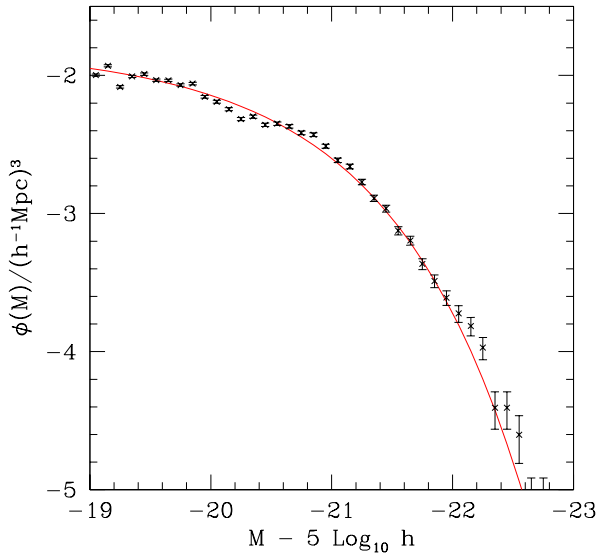} 
 \caption{Luminosity function in the mock catalog (symbols with 
          error bars); $M$ refers to the absolute magnitude in the 
          $r-$band.  Smooth curve shows the SDSS luminosity 
          function (Blanton et al. 2003).}
 \label{lf}
\end{figure}

The resulting mock galaxy catalog has been constructed to have the 
correct luminosity function (Figure~\ref{lf}) as well as the correct 
luminosity dependence of the galaxy two-point correlation function.  
In addition, note that the number of galaxies in a halo, 
the spatial distribution of galaxies within a halo, and the 
assignment of luminosities all depend only on halo mass, and not 
on the surrounding large-scale structure.  Therefore, the mock catalog 
includes only those environmental effects which arise from the 
environmental dependence of halo abundances.  

For reasons described by Sheth, Connolly \& Skibba (2005), 
the marked correlation function we measure in the mock catalogs is 
\begin{equation}
 M(s) \equiv {1 + W(s)\over 1 + \xi(s)},
 \label{markedXi}
\end{equation}
where $\xi(s)$ is the two-point correlation function in redshift 
space, and $W(s)$ is the same sum over galaxy pairs separated in 
redshift space by $s$, but now each member of the pair is weighted by 
the ratio of its luminosity to the mean luminosity of all the galaxies 
in the mock catalog.  (Schematically, if the estimator for $1+\xi$ 
is $DD/RR$, then the estimator for $1+W$ is $WW/RR$, so the 
estimator we use for $M$ is $WW/DD$.)
This measurement of $M(s)$ represents the prediction of the 
`standard' model:  the shape of the luminosity weighted correlation 
function includes the effects of the statistical correlation between 
halo mass and environment, but no other physical effects.  

\section{The halo model description}\label{model}
This section shows how to describe the marked correlation 
in redshift space discussed above in the language of the halo model.  
Details are in Skibba \& Sheth (2005); in essence, the calculation 
combines the results of Sheth (2005) with those of Seljak (2001).  

In the halo model, all mass is bound up in dark matter halos which 
have a range of masses.  Hence, the density of galaxies is 
\begin{equation}
  \bar n_{\rm gal} = \int dm \, {dn(m)\over dm}\, N_{\rm gal}(m),
\end{equation}
where $dn(m)/dm$ denotes the number density of haloes of mass $m$.  
The redshift space correlation function is the Fourier transform of 
the redshift space power spectrum $P(k)$:
\begin{equation}
 \xi(s) = \int {dk\over k}\, {k^3 P(k)\over 2\pi^2}\, {\sin ks\over 
ks}.
\end{equation}
In the halo model, $P(k)$ is written as the sum of two terms: 
one that arises from particles within the same halo and dominates 
on small scales (the 1-halo term), 
and the other from particles in different haloes which dominates 
on larger scales (the 2-halo term).  Namely, 
\begin{equation}
 P(k) = P_{1h}(k) + P_{2h}(k),
 \label{Pk1h2h}
\end{equation}
where, in redshift space,  
\begin{eqnarray}
 P_{1h}(k) &=& \int dm\,{dn(m)\over dm}\,
           \Biggl[{2N_{\rm sat}(m)\,u_1(k|m)\over
                  \bar n_{\rm gal}^2} \nonumber\\
   &&\qquad\qquad\qquad\qquad 
     +\ {N_{\rm sat}^2(m)\,u_2^2(k|m)\over
                  \bar n_{\rm gal}^2}\Biggr], \\
 P_{2h}(k) &=& \left(F_{\rm g}^2 + {2F_{\rm g}F_{\rm v}\over 3} 
                         + {F_{\rm v}^2\over 5}\right)\,P_{\rm 
Lin}(k),\\
 \label{u1}
 u_1(k|m) &=& \left[{\sqrt{\pi}\over 2}
          {{\rm erf}(k\sigma_{\rm vir}(m)/\sqrt{2}H)
            \over k\sigma_{\rm vir}(m)/\sqrt{2}H}\right]\,u(k|m),\\
 u^2_2(k|m) &=& \left[{\sqrt{\pi}\over 2}
                      {{\rm erf}(k\sigma_{\rm vir}(m)/H)
                       \over k\sigma_{\rm vir}(m)/H}\right]\,u^2(k|m),
 \label{u2}
\end{eqnarray}
$u(k|m)$ is the Fourier transform of the halo density profile divided 
by the mass $m$, $H$ is the Hubble constant, and 
\begin{equation}
 \sigma_{\rm vir}^2(m) \approx {Gm\over 2r_{\rm vir}} 
             = G\left({\pi\over 6} m^2 \Delta_{\rm 
vir}\bar\rho\right)^{1/3}
 \label{sigmavir}
\end{equation}
is the line-of-sight velocity dispersion within a halo 
($\Delta_{\rm vir}\approx 200$).  
In addition, the bias factor $b(m)$ describes the strength of halo 
clustering, 
\begin{eqnarray}
 F_{\rm v} &=& f \int {\rm d}m\,{{\rm d}n(m)\over{\rm d}m}\,
                 {m\over\bar\rho}\,u_1(k|m)\,b(m),\\
 F_{\rm g} &=& \int {\rm d}m\,{{\rm d}n(m)\over{\rm d}m}\,
                {1 + N_{\rm sat}(m)u_1(k|m)\over \bar n_{\rm 
gal}}\,b(m),
\end{eqnarray}
$f \equiv d\ln D(a)/d\ln a\approx \Omega^{0.6}$, and 
$P_{\rm Lin}(k)$ is the power spectrum of the mass in linear theory.  
The real space power spectrum is given by setting the terms in 
square brackets in equations~(\ref{u1}) and~(\ref{u2}) for 
$u_1$ and $u_2$ to unity, and $F_{\rm v}\to 0$.  
When explicit calculations are made, we assume that the density 
profiles of halos have the form described by Navarro et al. (1996), 
so $u$ has the form given by Scoccimarro et al. (2001), 
and that halo abundances and clustering are described by the 
parameterization of Sheth \& Tormen (1999).  

To describe the effect of weighting each galaxy by its luminosity, 
let $W(r)$ denote the weighted correlation function, and ${\cal W}(k)$ 
its Fourier transform.  Following Sheth, Abbas \& Skibba (2004) and 
Sheth (2005), we write this as the sum of two terms:  
\begin{equation}
 {\cal W}(k) = {\cal W}_{1h}(k) + {\cal W}_{2h}(k), 
 \label{Wk1h2h}
\end{equation}
where 
\begin{eqnarray*}
 {\cal W}_{1h}(k) &=& \int dm\,{dn(m)\over dm}\nonumber\\
   && \ \times\ \Biggl[
   {2L_{\rm cen}(m)\,\langle L|m,L_{\rm min}\rangle\,N_{\rm 
sat}(m)\,u_1(k|m)
         \over \bar n_{\rm gal}^2\ \bar L^2} \nonumber\\
 && \qquad\qquad + \ 
   {\langle L|m,L_{\rm min}\rangle^2 N_{\rm sat}^2(m)\,u_2^2(k|m)\over 
                \bar n_{\rm gal}^2\ \bar L^2}\Biggr], \nonumber\\
 {\cal W}_{2h}(k) &=& \left(F_{\rm w}^2 + {2F_{\rm w}F_{\rm v}\over 3} 
                         + {F_{\rm v}^2\over 5}\right)\,P_{\rm Lin}(k),
\end{eqnarray*}
with 
\begin{eqnarray}
 F_{\rm w} &=& \int {\rm d}m\,{{\rm d}n(m)\over{\rm 
d}m}\,b(m)\,\nonumber\\
        &&\quad\times\ 
          {L_{\rm cen}(m) + N_{\rm sat}(m)\langle L|m,L_{\rm 
min}\rangle 
           u_1(k|m)\over \bar n_{\rm gal}\ \bar L}
\end{eqnarray}
and
\begin{equation}
 \bar L = \int dm\, {dn(m)\over dm}\, 
               {L_{\rm cen}(m) + N_{\rm sat}(m)\,\langle L|m,L_{\rm 
min}\rangle
                                 \over\bar n_{\rm gal}}\,.
\end{equation}
Here $\bar L$ is the average luminosity, 
$L_{\rm cen}(m)$ is the luminosity of the galaxy at the centre of 
an $m$-halo, 
and $\langle L|m,L_{\rm min}\rangle$ is the average luminosity of 
satellite galaxies more luminous than $L_{\rm min}$ in $m$-haloes.  
Thus, the calculation requires an estimate of how the central 
and the average satellite luminosity depend on $m$.  
As we show below, both are given by the luminosity dependence of 
$\xi$ (i.e., equation~\ref{sdssNg}), so this halo model calculation 
of the weighted correlation function requires {\em no} additional 
information!  

The luminosity of the central galaxy is obtained by inverting the 
relation between $M_{\rm min}$ and $L$ (e.g., equation~\ref{MLapprox}).  
Obtaining an expression for the average luminosity of a satellite 
galaxy is more complicated.  Define 
\begin{eqnarray}
 P(>L|m,L_{\rm min})&\equiv& {N_{\rm sat}(>L|m)
                            \over N_{\rm sat}(>L_{\rm 
min}|m)}\nonumber\\
           &=& \int_L^\infty dL\,p(L|m,L_{\rm min}),
\end{eqnarray} 
where $N_{\rm sat}$ is given by equation~(\ref{sdssNg}).  
Then the mean luminosity of satellites in $m$ halos, 
\begin{equation}
 \langle L|m,L_{\rm min}\rangle = 
   \int_{L_{\rm min}}^\infty dL\,p(L|m,L_{\rm min})\,L,
\end{equation}
can be obtained from the fact that 
\begin{eqnarray}
 \int_{L_{\rm min}}^\infty dL' \, P(>L'|m,L_{\rm min})
 &=& \langle L|m,L_{\rm min}\rangle - L_{\rm min}.
\end{eqnarray}
This shows that if we add $L_{\rm min}$ to the quantity on the left 
hand side (which is given by integrating equation~\ref{sdssNg} over 
$L$), we will obtain the quantity we are after.  

\begin{figure}
 \centering
 \includegraphics[width=\hsize]{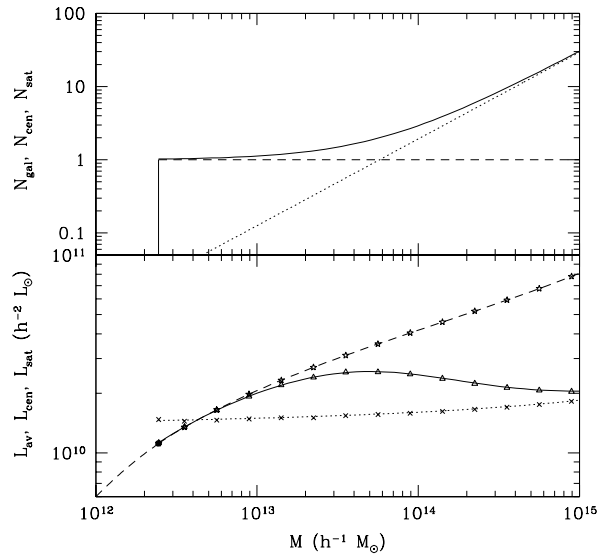} 
 \caption{Mean number of galaxies in a halo (top) and mean 
          luminosity in a halo (bottom) for SDSS galaxies with 
          $M_r<-20.5$, as a function of the masses of their parent 
          halos, predicted by the luminosity dependence of clustering.  
          Different curves in bottom panel show the mean luminosity 
          of the galaxies in a halo, the luminosity of the central 
          galaxy, and the mean luminosity of the others, as a 
          function of halo mass (solid, dashed, and dotted curves).
          Symbols show the result of computing these relations 
          in our mock catalogs.  }
 \label{compareML}
\end{figure}

Incidentally, since both $L_{\rm cen}(m)$ and 
$\langle L|m,L_{\rm min}\rangle$ can be estimated from the SDSS fits, 
the mean luminosity of the galaxies in an $m$-halo,
\begin{equation}
 L_{\rm av}(m,L_{\rm min}) = 
 {L_{\rm cen}(m) + N_{\rm sat}(m,L_{\rm min})\langle L|m,L_{\rm 
min}\rangle
 \over 1 + N_{\rm sat}(m,L_{\rm min})},
 \label{m2l}
\end{equation}
is completely determined by  equation~(\ref{sdssNg}).  
The mass-to-light ratio of an $m$ halo is 
$m/[N_{\rm gal}(m)\,L_{\rm av}(m)]$:  
this shows explicitly that the luminosity dependence of the galaxy 
correlation function constrains how the halo mass-to-light ratio must 
depend on halo mass.  This halo-mass dependence 
has been used by Tinker et al. (2005); our analysis provides an 
analytic calculation of the effect.  It shows that, in low mass halos, 
$L_{\rm av}\approx L_{\rm cen}$ because $N_{\rm sat}\ll 1$, 
whereas in massive halos, $L_{\rm av} < L_{\rm cen}$.  
Figure~\ref{compareML} compares the mass dependence of
$L_{\rm av}$, $L_{\rm cen}$, and $L_{\rm sat}$ for galaxies restricted
to $M_r<-20.5$ as predicted by Zehavi et al.'s (2005) halo model 
interpretation of the luminosity dependence of clustering in the SDSS.
The symbols show measurements from our mock catalogs.
The different quantities scale very differently 
with halo mass, with the following consequence.  

Equation~(\ref{Wk1h2h}) treats the central galaxies differently from 
the others.  If the luminosities of the central galaxies were not 
special, then the contribution to the one-halo term would scale as 
$N_{\rm sat}L_{\rm av}^2$ for the centre-satellite term, and 
$(N_{\rm sat}L_{\rm av})^2$ for the satellite-satellite term.  
Note that, in this case, the luminosity weights are the same for the 
two types of terms---only the number weighting differs.  However, 
because the mass dependence of $L_{\rm av}$ is so different from that 
of the other two terms (cf. Figure~\ref{compareML}), marked statistics 
allow one to discriminate between models which treat the central object 
as special from models which do not (e.g. Sheth 2005).  

\begin{figure}
 \centering
 \includegraphics[width=\hsize]{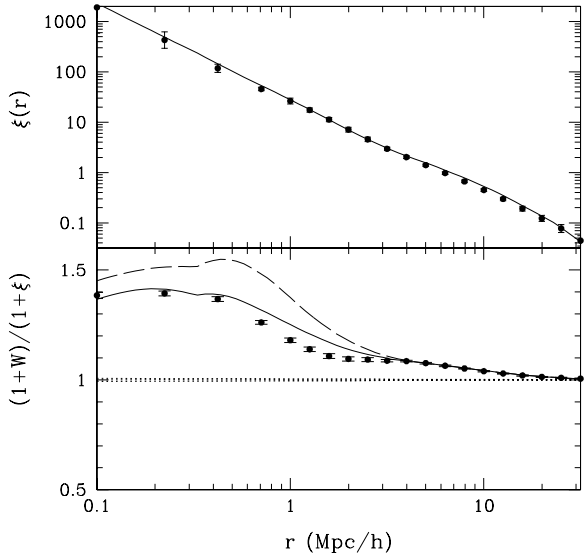} 
 \caption{Luminosity-weighted real-space correlation function 
          measured in a mock catalogs which resembles an SDSS 
          volume limited sample with $M_r<-20.5$.  Symbols 
          show the measurements; 
          smooth curves show the associated halo-model predictions 
          when the luminosity of the central galaxy in a halo is 
          assumed to be different from the others; dashed curves 
          show the prediction when the central object is not special.  
          Dotted curves show the mean and rms values of the 
          statistic $M$, obtained by after randomizing the marks and 
          remeasuring $M$ one hundred times.}
 \label{xirVLS}
\end{figure}

To illustrate, the symbols in the top and bottom panels of 
Figure~\ref{xirVLS} show measurements of $\xi(r)$ and $M(r)$ measured 
in this mock catalog.  
Error bars were obtained with a ``jack-knife'' procedure,
as detailed in Scranton et al. (2002),
in which the statistic is re-measured after omitting a random 
region, and repeating thirty times ($\sim$1.5 times the number of 
bins in separation for which we present results).  
Note that the errors in $W$ are strongly correlated with those in 
$\xi$, so that the true error in $M$ is grossly (more than a factor 
of ten) overestimated if one simply sums these individual errors in 
quadrature.  
A much better approximation of the uncertainties is obtained as 
follows.  Randomly scramble the marks among the galaxies, remeasure 
$M$, and repeat many ($\sim 100$) times.  
Compute the mean of $M$ over these realizations.  This mean, and 
the rms scatter around it are shown as dotted lines in the two 
panels.  Note that this scatter is within a factor of two of the 
full jackknife error estimate; it is smaller than the jackknife 
estimate on scales $r>1h^{-1}$Mpc and $s>3h^{-1}$Mpc, and, 
on smaller scales, it is larger than the jackknife estimate.  

The solid lines in the top panel show the halo model calculation 
of $\xi$.  These show that the model is in excellent agreement with 
the measurements on all scales in real space.  The solid and 
dashed curves in the bottom panel show the associated halo model 
calculations of the marked statistic $M$ when central galaxies are 
special (solid), and when they are not (dashed).  Note that both 
these curves give the {\em same} prediction for the unweighted 
statistic $\xi$.  

Comparison of these curves with the measurements yields two important 
pieces of information.  First, on large scales ($r>4h^{-1}$Mpc), the 
solid and dashed curves are identical, and they are in excellent 
agreement with the measurements.  
This indicates the large-scale signal is well described by a model 
in which there are no additional correlations with environment other 
than those which arise from the correlation between halo mass and 
environment.  This is not reassuring, since the mock catalogs were 
constructed to have no correlations other than those which are due 
to halo bias.  Second, on smaller scales, the solid curves are in 
substantially better agreement with the measurements than are the 
dashed curves.  (A $\chi^2$ estimate of the goodness of fit of the 
two marked correlation models yields values which smaller by a factor 
of ten when the central galaxy is treated specially compared to when 
it is not.)  Since the mock catalogs do treat the central galaxies 
differently from the others, it is reassuring that the halo model 
calculation which incorporates this difference is indeed in better 
agreement with the measurements.  

\begin{figure}
 \centering
 \includegraphics[width=\hsize]{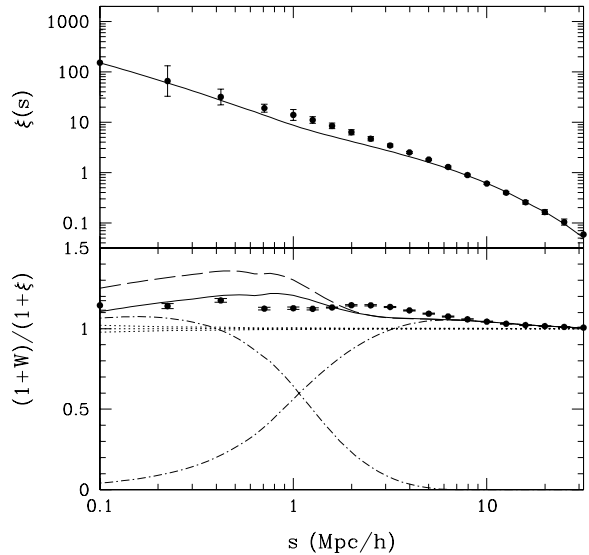} 
 \caption{Luminosity-weighted redshift-space correlation functions 
          measured in mock catalogs which resemble an SDSS 
          volume limited sample with $M_r<-20.5$.  Symbols 
          show the measurements;
          smooth curves show the associated halo-model predictions 
          when the luminosity of the central galaxy in a halo is 
          assumed to be different from the others; dashed curves 
          show the prediction when the central object is not special.  
          Dotted curves show the mean and rms values of the 
          statistic $M$, obtained by after randomizing the marks and 
          remeasuring $M$ one hundred times.  Dot-dashed curves in 
          show the one- and two-halo contributions to the statistic 
          in our model when the central object in a halo is special.}
 \label{xisVLS}
\end{figure}

In the next section, we will present measurements of marked 
statistics in redshift space.  To see if we can use our halo model 
calculation to interpret the measurements, Figure~\ref{xisVLS}
compares measurements of $\xi(s)$ and $M(s)$ in the mock catalog 
with our halo model calculation.  The format is similar to 
Figure~\ref{xirVLS}:  solid curves in the bottom panels show the 
predicted marked statistic $M$ when the central galaxy in a halo is 
treated differently from the others, and dashed curves show what 
happens if it is not.  Both curves give the same prediction for the 
unweighted statistic $\xi$.  

The top panel shows that the halo model calculation of $\xi(s)$ 
is in excellent agreement with the measurements on scales larger than 
a few Mpc, as it was for $\xi(r)$.  However, it is not as accurate 
when the redshift separations are of order a few Mpc.  Nevertheless, 
the model is able to reproduce the factor of ten difference between 
$\xi(r)$ and $\xi(s)$ on small scales.  We will discuss the reason 
for the discrepancy on intermediate scales shortly.  

Similarly, the bottom panel shows excellent agreement between 
measurements and model for the marked statistic $M(s)$ on large 
scales ($s>8h^{-1}$Mpc), both when the central object is treated 
specially and when it is not.  In addition, the model in which the 
central object is special is in better agreement with the measurements 
on small scales.  (A $\chi^2$ estimate of the 
goodness of fit of the two marked correlation models yields values 
which smaller by more than a factor of two when the central galaxy 
is treated specially compared to when it is not.)  
On intermediate scales, however, there is substantial discrepancy 
discrepancy between the model and the mocks; the discrepancy is 
more pronounced for $M(s)$ than for $\xi(s)$.

To study the cause of this discrepancy, dot-dashed lines show the 
two contributions to the statistic, ${\cal W}_{1h}/(1+\xi)$ and 
$(1+{\cal W}_{2h})/(1+\xi)$, separately.  This shows that it is 
on scales where both terms contribute that the model is inaccurate.  
There are two reasons why it is likely that this inaccuracy can be 
traced to our simple treatment of the two-halo term.  
The suppression of power due to virial motions means that we must 
model the two-halo term more accurately in redshift-space than in 
real-space.  Our halo-model calculation incorrectly assumes that 
linear theory is a good approximation even on small scales 
(e.g. Scoccimarro 2004 shows that this is a dangerous assumption 
even on scales of order 10 Mpc) and that volume exclusion effects 
(Mo \& White 1996) are negligible (Sheth \& Lemson 1999 discuss how 
one might incorporate such effects).  Because our mocks make use of 
both the positions and velocities of the halos in the simulations, 
they incorporate both these effects.  Thus, our simple halo-model 
likely underestimates $M(s)$ on intermediate scales, but overestimates 
it on smaller scales.  Since this is in the sense of the discrepancy 
with the measurements in the mock catalogs, it is likely that this 
inaccuracy can be traced to our simple treatment of the two-halo 
term.  
We will have cause to return to 
this discrepancy in the next section, where we use our halo model 
calculation to interpret measurements of marked statistics in 
the SDSS dataset.


\section{Measurements in the SDSS}\label{sdss}
\begin{figure}
 \centering
 \includegraphics[width=\hsize]{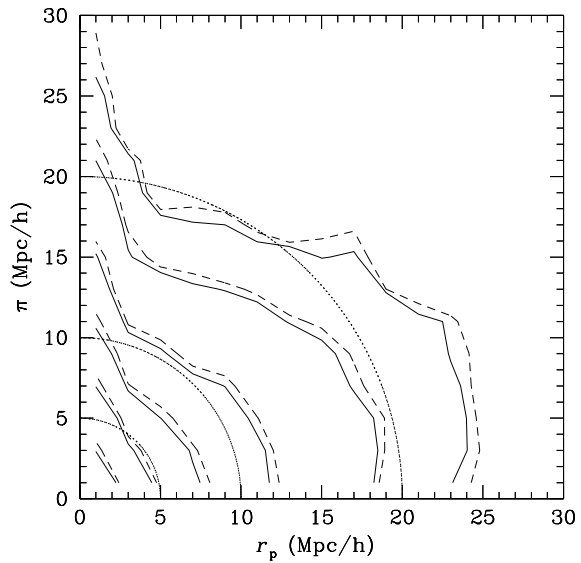} 
 \caption{Unweighted (solid) and weighted (dashed) correlation 
          functions measured in volume limited catalog with 
          $M_r<-20.5$ in the SDSS.  Dotted curves show that 
          the measured correlation functions are significantly 
          anisotropic.  }
 \label{rppiSDSS}
\end{figure}

Figure~\ref{rppiSDSS} shows $\xi(r_p,\pi)$ and $W(r_p,\pi)$, 
the unweighted (solid) and weighted (dashed) correlation functions 
of pairs with separations $r_p$ and $\pi$, perpendicular and 
parallel to the line of sight.  
The measurements were made in a volume limited catalog (59,293
galaxies with $M_r<-20.5$) extracted from the SDSS DR4 
database (Adelman-McCarthy et al. 2005).  
Contours show the scales at which the correlation functions 
have values of 0.1, 0.2, 0.5, 1, 2, and 5, when averaged over 
bins of $2h^{-1}$Mpc in $r_p$ and $\pi$.  This format, due to 
Davis \& Peebles (1983), allows one to isolate redshift space 
effects on the correlation functions, since these act only in 
the $\pi$ direction.  The dotted quarter circles at separations 
of 5, 10 and $20h^{-1}$Mpc are drawn to guide the eye---they serve 
to highlight the fact that both $\xi(r_p,\pi)$ and $W(r_p,\pi)$ are 
very anisotropic.  In constrast, the corresponding real-space 
quantities would be isotropic.  The figure shows clearly that $W$ 
has a slightly higher amplitude than $\xi$ on the scales shown.  

\begin{figure*}
 \centering
 \includegraphics[width=0.45\hsize]{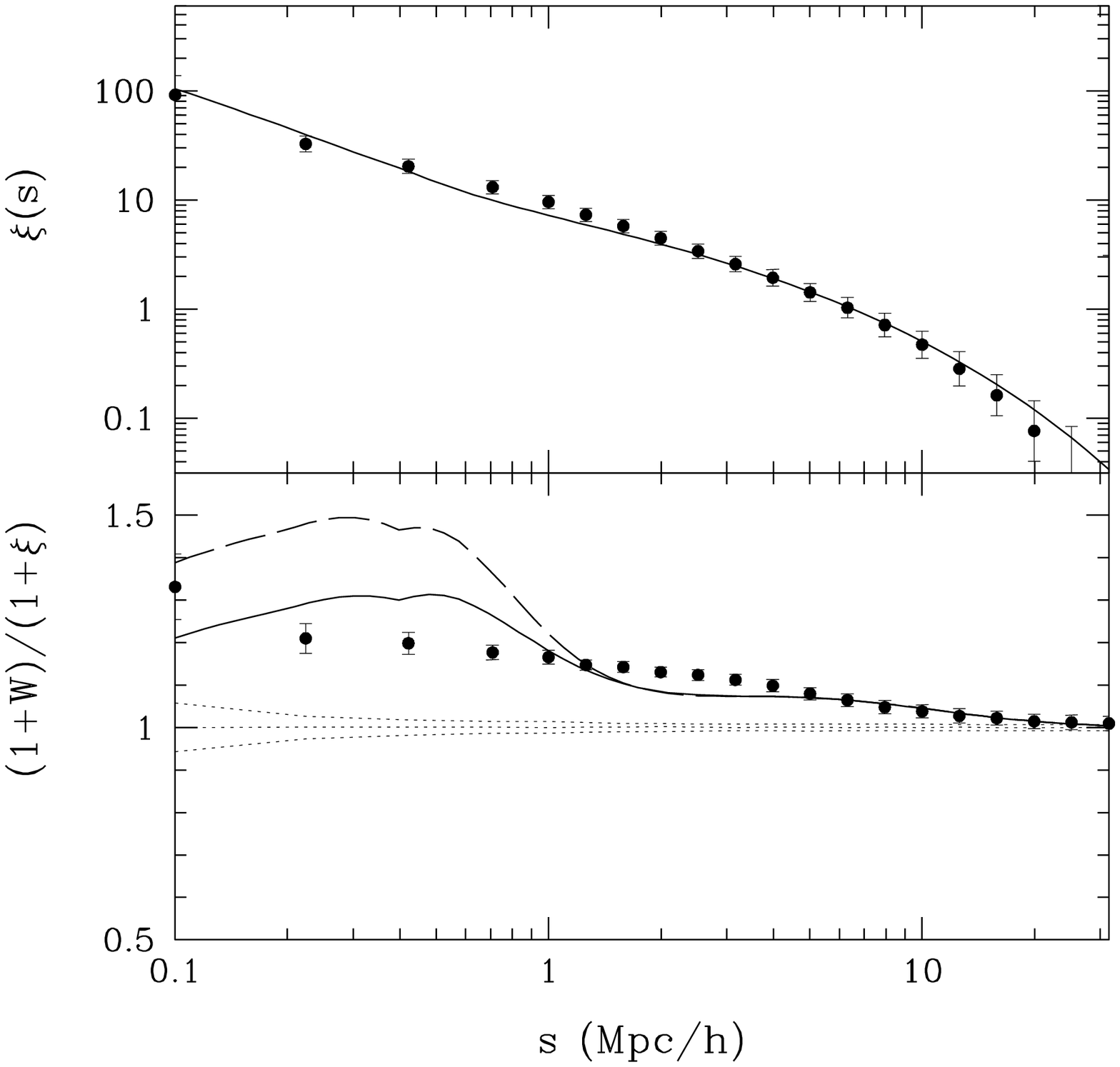} 
 \includegraphics[width=0.45\hsize]{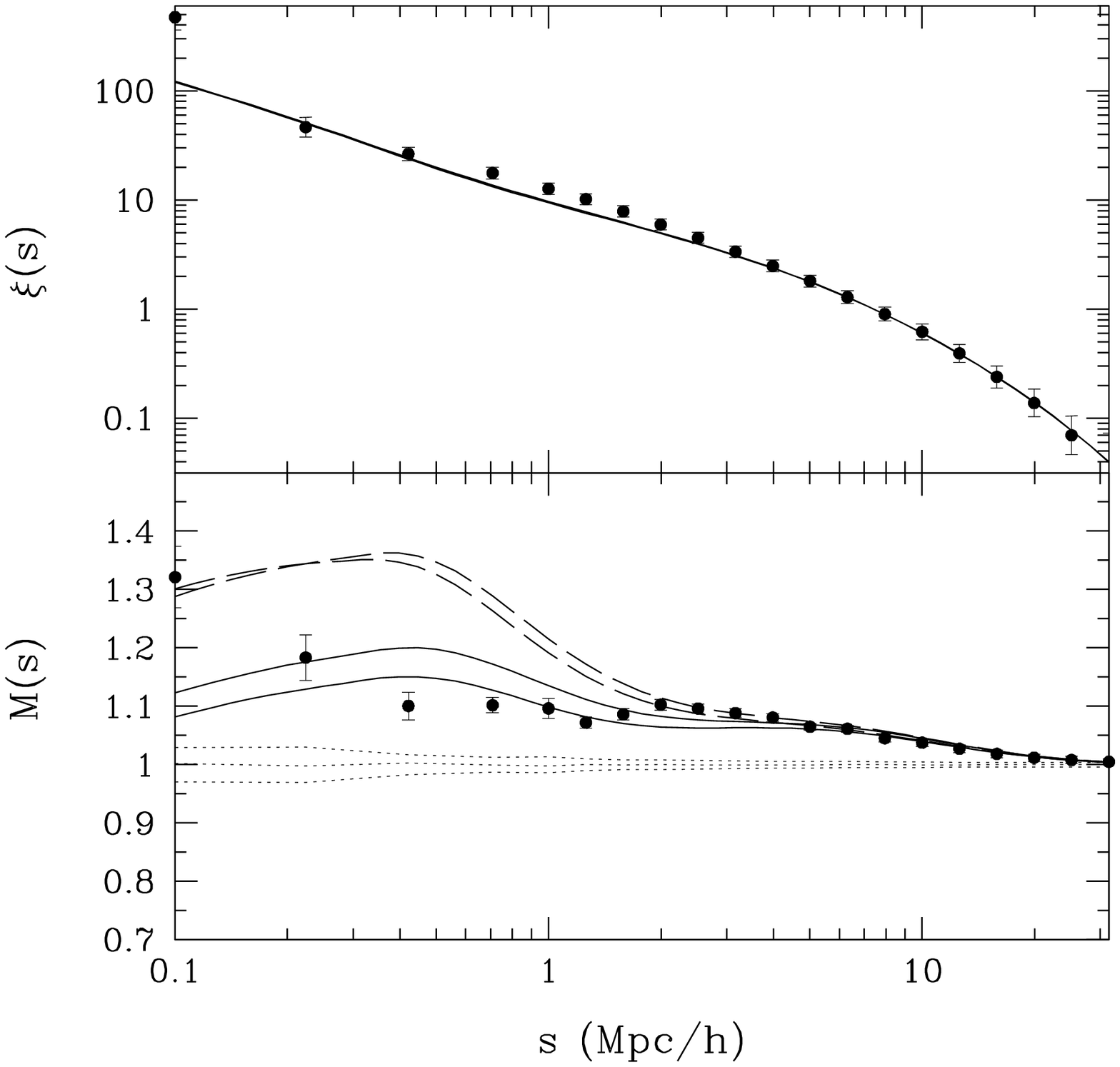} 
 \caption{Redshift-space correlation functions measured in volume 
          limited catalogs with $M_r<-19.5$ (left) and $M_r<-20.5$ 
          (right) in the SDSS.  
          Top panels show the unweighted correlation function $\xi(s)$, 
          and bottom panels show the marked statistic $M(s)$.  
          Smooth curves show the associated redshift-space 
          halo-model predictions; solid curves are when  
          the central galaxy in a halo is treated differently from 
          the others, whereas this is not done for the dashed curves.
          Dotted curves show the mean and rms values of the statistic 
          $M$, estimated by randomizing the marks and remeasuring $M$ 
          one hundred times.  Two sets of curves are shown in the right 
          hand panels; the top set of solid and dashed curves shows 
          the halo model calculation in which the relation between the 
          number of galaxies and halo mass is given by 
          equation~(\ref{sdssNg}), and the bottom set follow from 
          equation~(\ref{sdssNerfc}).}
 \label{xiSDSS}
\end{figure*}

\begin{figure}
 \centering
 \includegraphics[width=\hsize]{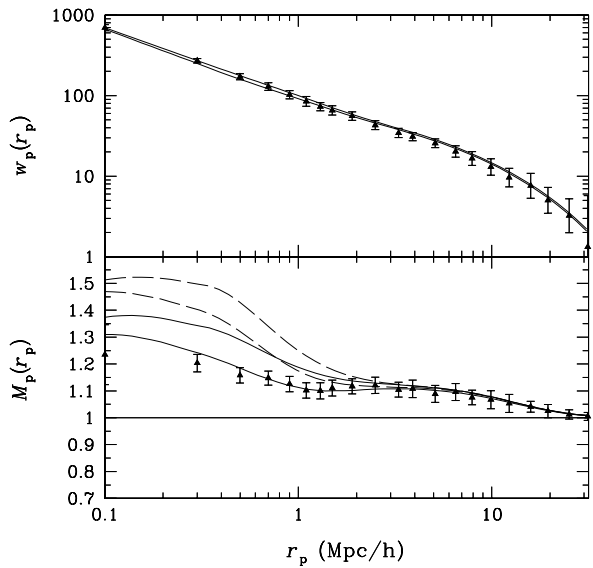} 
 \caption{Projected correlation function measured in volume limited 
          catalog with $M_r<-20.5$ in the SDSS.  Top panel shows the 
          unweighted projected correlation function $w_p(r_p)$,
          and bottom panels show the marked statistic $M_p(r_p)$.  
          Smooth curves show the associated projected halo-model 
          predictions; solid curves are when the central galaxy in a 
          halo is treated differently than the others, whereas this 
          is not done for the dashed curves.  The upper set of 
          dashed and solid curves show halo-model calculations 
          which follow from equation~(\ref{sdssNg}); the lower set 
          of curves assume equation~(\ref{sdssNerfc}).}
 \label{projMp}
\end{figure}

The quantity studied in the previous section, for which we have 
analytic (halo-model) estimates, can be derived from this plot as 
follows.  Counting pairs in spherical shells of radius 
 $s = \sqrt{r_p^2 + \pi^2}$ 
yields the redshift space correlation function $\xi(s)$.  
This measure of clustering is sensitive to the fact that the 
correlation function in redshift-space is anisotropic; in particular, 
it contains information about the typical motions of galaxies within 
halos (which are responsible for the elongation of the contours 
along the $\pi$ direction at $r_p\le 5h^{-1}$~Mpc), as well as the 
motions of the halos themselves (which are responsible for the 
squashing along the $\pi$ direction at $r_p\ge 5h^{-1}$~Mpc).  
The result of counting pairs of constant $r_p$, whatever their 
value of $\pi$, yields the projected correlation function $w_p(r_p)$; 
since $r_p$ is not affected by redshift space distortions, this 
quantity contains no information about galaxy or halo motions, so 
is more closely related to the real-space correlation function.  
Figures~\ref{xiSDSS} and~\ref{projMp} compare both these quantities 
with the corresponding halo-model calculations.  

Figure~\ref{xiSDSS} shows $\xi(s)$ and $M(s)$ measured in two 
volume limited catalogs extracted from the SDSS database.  One of 
these catalogs is the same as that which resulted in 
Figure~\ref{rppiSDSS}, and the other is for a slightly fainter 
sample ($M_r<-19.5$, with 61,821 galaxies).  
Error bars are estimated by jack-knife resampling, as discussed 
previously.  The solid curves show the redshift-space halo-model 
calculation in which central galaxies are special, and dashed curves 
show the expected signal if they are not.  (Recall that both have 
the same $\xi(s)$.)  

On large scales, both the solid and dashed curves provide an 
excellent description of the measurements on large scales.  
This agreement suggests that correlations with environment on scales 
larger than a few Mpc are {\em entirely a consequence of the 
correlation between halo abundances and environment}, just as they 
were in the mock catalogs.  Since the model calculation incorporates 
the assumption that the halo mass function is top-heavy in dense regions, 
the agreement with the measured $M(s)$ provides strong evidence that 
this is indeed the case.  

The discrepancy between the halo-model calculation and the 
measurements on intermediate scales is similar to the discrepancy 
between the halo-model and the mock catalogs studied in the 
previous section.  There we argued that this is almost certainly 
due to our simple treatment of the two-halo contribution to the 
statistic.  Indeed, the marked statistics in the mock catalogs 
behave qualitatively like those in the SDSS data (compare 
Figures~\ref{xisVLS} and~\ref{xiSDSS}), suggesting that the 
discrepancy between the halo-model calculation and the measurements 
are due to this, rather than to any environmental effects operating 
on intermediate scales.  

On small scales, the solid curves are in substantially better 
agreement with the data than are the dashed curves ($\chi^2$ 
smaller by a factor of four in both plots).  
Evidently, central galaxies are indeed a special population in 
the data.  This provides substantial support for the assumption 
commonly made in halo-model interpretations of the galaxy correlation 
function that the central galaxy in a halo is different from all the 
others.

However, even the solid curves are not in particularly good 
agreement with the measurements.  Before attributing the discrepancy 
to environmental effects not included in the halo model description, 
we have explored the effect of modifying our parametrization of the 
relation between the number of galaxies and halo mass which we use 
(equation~\ref{sdssNg}).  Figure~\ref{xiSDSS} shows that the 
parametrization in equation~(\ref{sdssNerfc}), with $\sigma=0.5$ 
and $M_1/M_{\rm min}=30$, provides equally good fits to $\xi(s)$, 
but a slightly better description of $M(s)$.  In this parameterization 
of the scaling of $N_{\rm gal}$ with halo mass, the minimum halo 
mass required to host a galaxy is not a sharp step function.  

Further evidence in support of the parametrization in which the 
minimum mass cutoff is not sharp, and in which the central galaxy is 
different from the others is shown in Figure~\ref{projMp}.  
The top and bottom panels compare measurements of the projected 
correlation functions $w_p(r_p)$ and $M_p(r_p)$, where 
\begin{eqnarray}
 w_p(r_p) &=& \int {\rm d}y\,\xi(r_p,y) 
      = 2\int_{r_p}^\infty {\rm d}r\,{r\,\xi(r)\over\sqrt{r^2-r_p^2}},\quad
        {\rm and}\nonumber\\
 M_p(r_p) &=& {1+W_p(r_p)/r_p\over 1 + w_p(r_p)/r_p},
      \qquad {\rm where}\nonumber \\
 W_p(r_p) &=& 2\int_{r_p}^\infty {\rm d}r\,{r\,W(r)\over\sqrt{r^2-r_p^2}}
        \ {\rm and}\ r = \sqrt{r_p^2+y^2},
\end{eqnarray}
with the associated halo-model calculations.  
(In the halo model, the real-space quantities $\xi(r)$ and $W(r)$ 
which appear in the expressions above, are related to $\xi(s)$ and 
$W(s)$ by setting setting $F_{\rm v}=0$ and taking the limit 
$\sigma_{\rm vir}\to 0$ in $u_1$ and $u_2$.  See Skibba \& Sheth 2006 
for our particular definition of $M_p$.)  
Note that these projected quantities are free of redshift space 
distortions, making them somewhat easier to interpret.  

As was the case for the redshift space measurements, both 
parameterizations of $N_{\rm gal}(M)$ provide good descriptions of 
the unweighted statistic $w_p(r_p)$, and in both cases, the 
weighted statistic is in better agreement when the central 
object is treated specially.  However, the Figure shows clearly 
that when the central object is special, then 
equation~(\ref{sdssNerfc}) provides a substantially better 
description of $M_p$---the agreement with the measurements is 
excellent over all scales.  


\section{Discussion}
We showed how to generate a mock galaxy catalog which has the same 
luminosity function (Figure~\ref{lf}) and luminosity dependent 
two-point correlation function as the SDSS data.  
We used the mock catalog to calculate the luminosity-weighted 
correlation function in a model where all environmental effects are 
a consequence of the correlation between halo mass and environment 
(Figures~\ref{xirVLS} and~\ref{xisVLS} show results in real and 
redshift space).  
We then showed how to describe this luminosity-weighted correlation 
function in the language of the redshift-space halo model 
(equation~\ref{Wk1h2h}).  
The analysis showed that estimates of the luminosity dependence of 
clustering constrain how the mass-to-light ratio of halos depends on 
halo mass (equation~\ref{m2l} and Figure~\ref{compareML}).  
The central galaxy in a halo is predicted to be substantially 
brighter than the other objects in the halo, and although the 
luminosity of the central object increases rapidly with halo 
mass, the mean luminosity of the other objects in the halo is 
approximately independent of the mass of the host halo.  

Our analysis also showed that measurements of clustering as a 
function of luminosity completely determine the simplest halo model 
description of marked statistics.  In addition, measurements of the 
marked correlation function allow one to discriminate between models 
which treat the central object in a halo as special, from those which 
do not (Figures~\ref{xirVLS} and~\ref{xisVLS}).  
Also, in hierarchical galaxy formation models, the marked 
correlation function is expected to show a signal on large scales 
if the average mark of the galaxies in a halo correlates with halo 
mass.  
The signal arises because massive halos populate the densest regions; 
it is present even if there are no physical effects which operate to 
correlate the marks over large scales.  

We compared this halo model of marked statistics with measurements 
in the SDSS (Figures~\ref{xiSDSS} and~\ref{projMp}).  
The agreement between the model and the measurements on scales 
smaller than a few Mpc provides strong evidence that central galaxies 
in halos are a special population---in general, the central galaxy 
in a halo is substantially brighter than the others.  
(Berlind et al. 2004 come to qualitatively similar conclusions, 
but from a very different approach.)  Substantially better agreement 
is found for a model in which the minimum halo mass required to host 
a luminous central galaxy does not change abruptly with luminosity.  
This is in qualitative agreement with some semi-analytic galaxy 
formation models, which generally predict some scatter in central 
luminosity at fixed halo mass (e.g. Sheth \& Diaferio 2001; 
Zheng et al. 2005).  

The agreement between the halo model calculation and the data on 
scales larger than a few Mpc indicates that the standard assumption 
in galaxy formation models, that halo mass is the primary driver of 
correlations between galaxy luminosity and environment, is accurate.  
In particular, these measurements are consistent with a model in 
which the halo mass function in large dense regions is top-heavy, 
and, on these large scales, there are no additional physical or 
statistical effects which affect the luminosities of galaxies.  
In this respect, our conclusions are similar to those of 
Mo et al. (2004), Kauffmann et al. (2004), Blanton et al. (2005), 
and Abbas \& Sheth (2006), although our methods are very different.  

We note in passing that there is a weak statistical effect for which 
the halo-model above does not account:  at fixed mass, haloes in dense 
regions form earlier (Sheth \& Tormen 2004).  
Gao, Springel \& White (2005) show that this effect is more pronounced 
for low mass haloes (related marked correlation function analyses by 
Harker et al 2006 and Wechsler et al. 2006 come to similar conclusions).  
The agreement between our halo-model calculation and the measurements 
in the SDSS suggests that this correlation between halo formation and 
environment is not important for the relatively bright galaxy 
population we have studied here.  This is presumably because these 
SDSS galaxies populate more massive haloes.  Comparisons with larger 
upcoming SDSS datasets, with a fainter luminosity threshold 
(such as $M_r<-18$), may bear out the correlation between low-mass halo 
formation and environment.  

\begin{figure}
 \centering
 \includegraphics[width=\hsize]{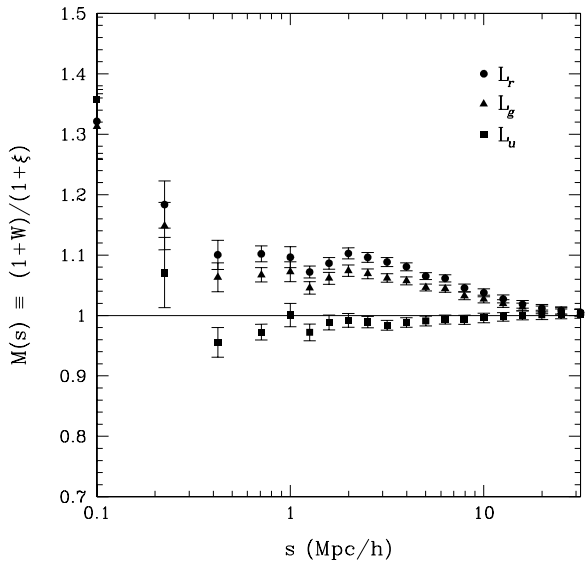} 
 \caption{Redshift-space luminosity-weighted correlation functions 
          measured in volume limited catalogs with $M_r<-20.5$ in 
          the SDSS.  
          Circles, triangles and squares show $M(s)$ when the weight 
          is $r-$, $g-$ and $u-$band luminosity respectively.  
          For clarity, jack-knife error bars are only shown for the 
          $r-$band measurement, since the uncertainties are similar 
          in the other bands.}
 \label{ugrMs}
\end{figure}

As a final indication of the information contained in measurements 
of marked statistics, Figure~\ref{ugrMs} compares $M(s)$ when the 
$u-$, $g-$ and $r-$band luminosities are used as the mark.  The 
underlying population is the same as that for 
Figures~\ref{rppiSDSS}--\ref{projMp}:  the sample is volume limited 
to $M_r<-20.5$.  
Thus, $\xi(s)$ is fixed, and only $W(s)$ changes with wave-band.  
Notice that there is a clear trend with wavelength:  there is 
a slight anti-correlation 
in the $u-$band, whereas $M(s)$ rises slightly with 
decreasing scale when $g-$band luminosity is the mark, 
and the signal is even stronger when $L_r$ is the mark.  
This trend with wavelength is qualitatively consistent with the 
predictions of semi-analytic galaxy formation models (Sheth, 
Connolly \& Skibba 2005) and indicates that the mean $u$-band 
luminosity of the galaxies in a halo depends less strongly on halo 
mass than does the mean $r-$band luminosity in a halo (Sheth 2005).  
In the models, the $u-$band luminosity is an indicator of the star 
formation rate; our measurements suggest that the correlation between 
star formation rate and halo mass is weak---if it is an increasing 
function of halo mass at low masses, then it decreases at larger 
masses.  

\begin{figure}
 \centering
 \includegraphics[width=\hsize]{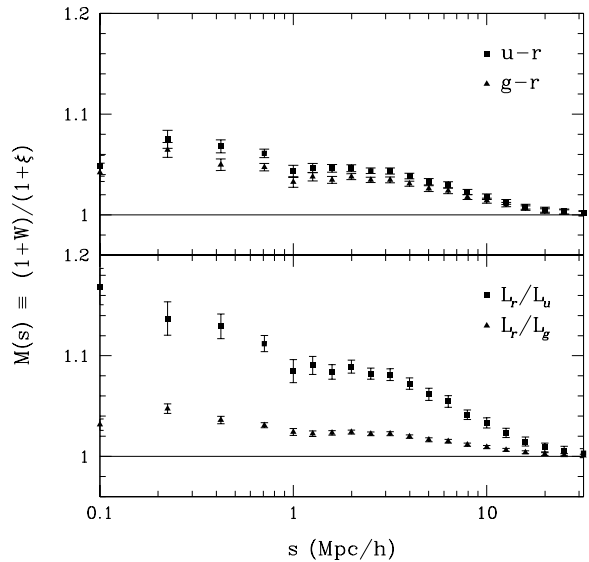} 
 \caption{Redshift-space color-weighted correlation functions 
          measured in volume limited catalogs with $M_r<-20.5$ in 
          the SDSS.  The top panel shows results when the color weight 
          is the difference in absolute magnitudes.  In the bottom 
          panel, galaxies were weighted by the ratio of the 
          luminosities in the two bands, so they span a greater range 
          around the mean value.  
          Both panels show that close pairs of galaxies tend to have 
          redder colors, although the difference is clearer when the 
          weights span a greater range around the mean value.}
 \label{umrgmr}
\end{figure}

Figure~\ref{umrgmr} shows the result of weighting these same galaxies 
by their colors.  The top panel shows results where the weight is the 
difference in the absolute magnitudes, $M_u-M_r$ and $M_g-M_r$, 
whereas the weights in the bottom bottom panel were the ratios of the 
luminosities in two bands.  Comparison of the two panels shows the 
effect on $M(s)$ of rescaling the weights while preserving their 
rank-ordering---while there are quantitative differences, the 
results in both panels are qualitatively similar.  The $M(s)$ 
measurements shown in the bottom panel are more widely separated 
because the luminosity ratio involves $10^{\rm color}$, which has 
the effect of weighting the redder galaxies more heavily. 
In particular, this analysis indicates clearly that close pairs of 
galaxies tend to be redder than average.  
Sheth, Connolly \& Skibba (2005) show that this is also the case in 
semi-analytic galaxy formation models.  

The measurements shown in Figures~\ref{ugrMs} and~\ref{umrgmr} are 
consistent with models in which galaxies in clusters are more massive 
and have smaller star formation rates than galaxies in the field.  
In effect, these figures demonstrate the environmental dependence of 
galaxy luminosities and colors, without having to divide the galaxy 
sample up into discrete bins of `field', `group', and `cluster'.  
Thus, marked statistics allow one to study correlations with 
environment 
over a continuous range in density, rather than in somewhat arbitrary 
discrete bins in environment.  In this respect, our use of marked 
statistics to quantify and interpret environmental trends is very 
different from recent approaches which address the same problem.  
Since marked statistics are simple to measure and interpret, we 
hope that they will become standard tools for quantifying the 
correlation between the properties of galaxies and their environments.  

\section*{Acknowledgements}
We thank the Virgo consortium for making their simulations available 
to the public ({\tt www.mpa-garching.mpg.de/Virgo}), 
and the Pittsburgh Computational Astrostatistics group (PiCA) for 
the NPT code which was used to measure the correlation functions 
presented here.
This work was supported by NASA-ATP NAG-13720 and by the NSF under 
grants AST-0307747 and AST-0520647.  

Funding for the SDSS and SDSS-II has been provided by the 
Alfred P. Sloan Foundation, the Participating Institutions, 
the National Science Foundation, the U.S. Department of Energy, 
the National Aeronautics and Space Administration, the Japanese 
Monbukagakusho, the Max Planck Society, and the Higher Education 
Funding Council for England. The SDSS Web Site is http://www.sdss.org/.

The SDSS is managed by the Astrophysical Research Consortium for 
the Participating Institutions. The Participating Institutions are 
the American Museum of Natural History, Astrophysical Institute 
Potsdam, University of Basel, Cambridge University, Case Western 
Reserve University, University of Chicago, Drexel University, 
Fermilab, the Institute for Advanced Study, the Japan Participation 
Group, Johns Hopkins University, the Joint Institute for Nuclear 
Astrophysics, the Kavli Institute for Particle Astrophysics and 
Cosmology, the Korean Scientist Group, the Chinese Academy of 
Sciences (LAMOST), Los Alamos National Laboratory, the 
Max-Planck-Institute for Astronomy (MPA), the Max-Planck-Institute 
for Astrophysics (MPIA), New Mexico State University, Ohio State 
University, University of Pittsburgh, University of Portsmouth, 
Princeton University, the United States Naval Observatory, and 
the University of Washington.

\label{lastpage}

\end{document}